\documentclass[final,3p,times,twocolumn]{elsarticle}
\usepackage{fleqn,epsf}
\input{pstricks}
\biboptions{comma,sort&compress}
\usepackage{here}
\usepackage{epsf}
\usepackage{graphicx}
\newcommand{\be}{\begin{equation}}
\newcommand{\ee}{\end{equation}}
\newcommand{\bea}{\begin{eqnarray}}
\newcommand{\eea}{\end{eqnarray}}

\newcommand{\AmS}{{\protect\the\textfont2
  A\kern-.1667em\lower.5ex\hbox{M}\kern-.125emS}} 
\journal{Nuc. Phys. (Proc. Suppl.)}
\begin{document}
\title{Test of $SU(3)$ Symmetry in Hyperon Semileptonic
  Decays\tnoteref{t1}}
\tnotetext[t1]{Talk given at QCD14, the 17th International QCD Conference, 30 June--4 July 2014, Montpellier (France).}
\author[label]{T. N. Pham}
\address[label]{Centre de Physique Th\'eorique,
CNRS, Ecole Polytechnique, 91128 Palaiseau Cedex, France}   

\begin{abstract}
Existing analyzes  of baryon semileptonic decays indicate the presence
of a small $SU(3)$ symmetry breaking in hyperon semileptonic decays, but 
to provide evidence for $SU(3)$ symmetry breaking, one would need a relation
similar to the Gell-Mann--Okubo (GMO) baryon mass formula which is satisfied
to a few percents, showing  evidence for a small $SU(3)$ symmetry breaking
effect in the GMO mass formula. In this talk, I would like to  present
a similar GMO relation obtained in a recent work 
for hyperon semileptonic decay axial vector current matrix elements. 
Using these generalized  GMO relations for  the measured axial vector
 current to vector current form factor ratios, it is shown that  $SU(3)$
symmetry  breaking in hyperon semileptonic decays is of $5-11\%$ and
confirms the validity of the Cabibbo model for hyperon semi-leptonic decays.

\end{abstract}
\journal{Nuc. Phys. (Proc. Suppl.)}
\maketitle
\section{Introduction}
     The success of the Gell-Mann--Okubo (GMO) mass
 formula shows that $SU(3)$ is a good symmetry for strong interactions.
This approximate symmetry can be incorporated into a  QCD 
Lagrangian with $ m_{u},m_{d}\ll m_{s}$,with $m_{s}\ll \Lambda_{\rm QCD}$.
At low energies, an effective chiral Lagrangian can be constructed 
with baryons coupled to the pseudo-scalar meson octet, the 
Goldstone-Nambu boson of the $SU(3)\times SU(3)$ chiral symmetry.This 
Lagrangian contains the axial vector current matrix elements and 
produces the axial vector form factors measured in baryon semileptonic 
decays. The  Goldberger-Treiman  relation  for the pion-nucleon coupling
 constant is also obtained 
directly from this chiral Lagrangian. At zero order in the current
 $s$-quark $m_{s}$, the axial vector current form factors and the 
pseudo-scalar meson-baryon couplings are
 $SU(3)$-symmetric  and are completely given by the two parameters
$F$ and $D$ of the $F$(antisymmetric) and $D$(symmetric) type 
coupling~\cite{Donoghue} (in the standard chiral perturbation theory notation),
\bea
&&\kern -1.0cm {\cal L} = {\rm Tr} \biggl(\bar{B}(i{\not\cal D} -m_{0})B -
 D(\bar{B}\gamma_{\mu}\{\bar{\cal A}_{\mu},B\})\nonumber\\
&&- F(\bar{B}\gamma_{\mu}[\bar{\cal A}_{\mu},B])\biggr) + {\cal L}_{\rm SB}\\
&&\kern -1.0cm {\cal D}_{\mu}  = \partial_{\mu} + i\bar{\cal V}_{\mu},
\label{L}
\eea
with 

\bea
&&\kern -1.0cm \bar{\cal V}_{\mu}=-\frac{i}{2}\biggl( \xi^{\dagger}\partial_{\mu}\xi + \xi\partial_{\mu}\xi^{\dagger}\biggr), \nonumber\\
&& \kern -1.0cm \bar{\cal A}_{\mu}=-\frac{i}{2}\biggl( \xi^{\dagger}\partial_{\mu}\xi - \xi\partial_{\mu}\xi^{\dagger}\biggr),  \nonumber
\eea

  Assuming  $SU(3)$   symmetry, the  axial vector currents matrix
elements for  semileptonic hyperon decays can then  be described
by the two $SU(3)$-symmetric  $F$ and $D$ parameters as in the 
Cabibbo model~\cite{Cabibbo} for which the agreement with experiments
is quite good. The Cabibbo model for hadronic current for weak
interaction  semileptonic decays of hyperons is given as:
\be
 J_{\mu}= aJ_{\mu}^{0} + bJ_{\mu}^{1}
\label{model}
\ee 
where $J_{\mu}^{0}$  and $J_{\mu}^{1} $ are respectively, the two
$SU(3)$ octet  $\Delta S=0$ and $\Delta S=1$ left-handed hadronic currents.

    To have generalized universality for weak interaction,
Cabibbo assumes that the current $J_{\mu} $ has ``unit length''
and puts $a^{2} + b^{2}=1$, with 
$a=\cos{\theta}, b=\sin{\theta}$~\cite{Cabibbo}. This is exactly the
unitarity relation for the two-generation CKM quark mixing matrix,
\be
        |V_{ud}|^{2} + |V_{us}|^{2} = 1
\label{ckm2} 
\ee

    The remarkable fact is that, using $SU(3)$
symmetry for the matrix elements of $K$ and $\pi$ meson and hyperon
semileptonic decays, the predictions of the Cabibbo model are in good
agreement with experiments~\cite{Marshak}. The suppressed rates 
for  $K$ meson and hyperon semileptonic decays are explained. Since 
the Fermi coupling constant for nuclear $\beta$ decay is given 
by $G_{\beta}= G\cos{\theta}$, the discrepancy between $^{14}O$ and 
muon lifetime is resolved.

    Looking back, the Cabibbo model is a first step
toward the Cabibbo--Kobayashi--Maskawa (CKM) model for weak interactions,
 following the discovery of charm and bottom hadrons with the GIM 
mechanism and  CP-violation incorporated, with the CKM quark mixing
matrix:
\be
V_{\rm CKM} = \pmatrix{V_{ud}& V_{us} & V_{ub} \cr
V_{cd} & V_{cs} & V_{cb} \cr 
V_{td} & V_{ts} & V_{tb}} 
\label{Vckm} 
\ee
which can be parametrized by three mixing angles and the CP-violating KM
phase~\cite{Kobayashi}.

    Thus the success of the Cabibbo model comes out
 naturally with the GIM mechanism and the unitarity of the CKM matrix.
This model also confirms the validity of $SU(3)$ 
symmetry in hyperon semileptonic decays, but one expects some 
symmetry breaking effects.

    To have a precise determination of $V_{us}$
from hyperon semileptonic decays, one would need a good estimate of
 $SU(3)$ symmetry breaking effects. In the past there have been many 
works on  $SU(3)$ symmetry breaking in hyperon semileptonic decays, for 
example, in Refs.~\cite{Donoghue2, Roos, Gensini, Ratcliffe, Song, Dai,
  Manohar, Yamanishi, Holstein}  etc.  and more  recently
in Ref.~\cite{Batra}. These  works indicate the presence of a small $SU(3)$ 
symmetry breaking in hyperon semileptonic decays. But to provide
evidence for $SU(3)$ symmetry breaking, one needs
 a relation similar to the Gell-Mann--Okubo (GMO) baryon mass formula. In 
this talk I would like to present  GMO relations for hyperon semileptonic 
decays obtained in a recent work~\cite{Pham} and show that the amount of
$SU(3)$ symmetry breaking in  hyperon semileptonic decays is  $5-11\%$.

{\scriptsize
\begin{table*}[ht]
\begin{center}
\begin{tabular}{|c|c|c|c|c|}
\hline
\hline
 Decay &$f_{1}$ & $(g_{1}/f_{1})_{SU(3)+ \rm SB}$&
 $(g_{1}/f_{1})_{\rm exp}$\cite{PDG,Bourquin2} & $d_{B\to B^{'}}$(\rm est.) \\ 
\hline
$n\to p\ell\bar{\nu}$ &$1$ &$ F + D$&$1.2694\pm 0.0028 $&$$\\
$\Lambda\to p\ell\bar\nu$&$-\sqrt{3/2}$ & $F +
 D/3 + d_{\Lambda \to p}$&$0.718\pm 0.015 $&$ -0.015- 0.011 $\\
$\Sigma^-\to n\ell\bar\nu$&$-1$ & $F -D + d_{\Sigma^{-}\to
 n}$&$-0.340\pm 0.017 $&$ -0.034 $(\rm input) \\
$\Xi^-\to\Lambda^0\ell\bar\nu$ &$\sqrt{3/2}$ & $F
 -D/3+d_{\Xi^{-}\to \Lambda}$&$0.25\pm 0.05$&$0.053 - 0.023$ \\
$\Xi^0\to\Sigma^+\ell\bar\nu$ &$1$ &$F + D+d_{\Xi^{0}\to
 \Sigma^{+}} $&$1.21\pm 0.05 $&$-0.06$(\rm data)\\
$\Xi^-\to\Sigma^0\ell\bar\nu$&$1/\sqrt{2}$ &$F +
 D + d_{\Xi^{-}\to \Sigma^{0}}$&$ $&$ $\\
$\Sigma^-\to\Lambda\ell\bar\nu$ &$0$ &$(g_{1})_{SU(3)}=\sqrt{2/3}D $&$(g_{1})_{\rm exp}=$&$-0.070 - 0.028$\\
$\Sigma^+\to\Lambda\ell\bar\nu$ &$0$ & $(g_{1})_{SU(3)}=\sqrt{2/3}D $& $0.587\pm 0.016$&$-0.070 - 0.028$ \\
\hline
\hline
\end{tabular}
\end{center}
\caption{
Vector and axial vector current form factors for baryon 
semileptonic decays in the Cabibbo  model and with $SU(3)$ breaking 
term $d_{B\to B^{'}}$ and the measured axial vector to vector form 
factor ratio $g_{1}/f_{1}$, the $SU(3)$ and measured values for
 $(g_{1})_{\Sigma^-\to\Lambda}$. The last column is the estimated
$d_{B\to B^{'}} $.}
\label{table1}
\end{table*}}
\section{The GMO relations for hyperon semileptonic decays}

     In the standard model, $SU(3)$ symmetry breaking
 is given by the current quark mass term in the QCD Lagrangian 
with $m_{u,d} \ll m_{s}$. As an example, to derive the GMO relation
for baryon mass difference, consider now the  divergence of the
 $\Delta S=1$  V-spin $V=1$ vector current $\bar{u}\,\gamma_{\mu}\,s$. It 
is given by:
\be
\partial_{\mu}(\bar{u}\,\gamma_{\mu}\,s) = -i\,m_{s}\bar{u}\,s
\label{div}
\ee

    The  baryon mass difference 
is given by the $\bar{u}\,s $ scalar current form factor at the momentum
transfer $q=0$, in the limit of $SU(3)$ symmetry:
\bea
&&\kern-1.5cm  <\frac{1}{2}\Sigma^{0} +\frac{\sqrt{3}}{2}\Lambda|\bar{u}\,s|\Xi^{-}>= 
-<p|\bar{u}\,s|\frac{1}{2}\Sigma^{0} +\frac{\sqrt{3}}{2}\Lambda>\label{su1}\\
&&\kern-1.5cm  <\frac{\sqrt{3}}{2}\Sigma^{0} -\frac{1}{2}\Lambda|\bar{u}\,s|\Xi^{-}>= 
<p|\bar{u}\,s|\frac{\sqrt{3}}{2}\Sigma^{0} -\frac{1}{2}\Lambda >
\label{su2}
\eea
with  $|\Xi^{-}> = |V=1,V_{3} =1>$, $|p> = |V=1,V_{3} = -1>$ and
 $|\frac{1}{2}\Sigma^{0} +\frac{\sqrt{3}}{2}\Lambda> = |V=1,V_{3}=0>$,
 $|\frac{\sqrt{3}}{2}\Sigma^{0} -\frac{1}{2}\Lambda> = |V=0,V_{3}=0>$. 

 Eqs. (\ref{su1}--\ref{su2})  are  the rotated V-spin version  of the
 two I-spin relations for the   $\bar{u}\,d$ matrix elements:\break
    $ <B'|\partial_{\mu}(\bar{u}\,\gamma_{\mu}\,s)|B>$ given by
$ (f_{1})_{B\to B'}(m_{B'} -   m_{B})$,  $(f_{1})_{B\to B'}$ being
the  form factor at $q^{2}=0$ momentum transfer in  the vector current 
  $<B'|\bar{u}\,\gamma_{\mu}\,s|B>$ matrix element which has
no first order $SU(3)$ breaking effect according to the Ademollo-Gatto
theorem~\cite{Ademollo}. We thus have,
\bea
&&\kern-1.5cm  [(1/4)(m_{\Xi^{-}} -m_{\Sigma^{0}}) +(3/4)(m_{\Xi^{-}} -m_{\Lambda})]\nonumber\\
&&=[(1/4)(m_{\Sigma^{0}}-m_{p}) +(3/4)(m_{\Lambda}-m_{p})]
\label{GMO1}\\
&&\kern-1.5cm [(m_{\Xi^{-}} -m_{\Sigma^{0}}) -(m_{\Xi^{-}} -m_{\Lambda}))\nonumber\\
 &&=-[(m_{\Sigma^{0}}-m_{p}) -(m_{\Lambda}-m_{p})]
\label{GMO2}
\eea
   Eq. (\ref{GMO1})  reproduces  the GMO relation:
\bea
&&\kern-1.5cm  (3/4)(m_{\Lambda}-m_{N})+  (1/4)(m_{\Sigma} - m_{N})\nonumber\\
 &&= (3/4)(m_{\Xi} - m_{\Lambda} )+  (1/4)(m_{\Xi} - m_{\Sigma}) 
\label{GMO}
\eea
while Eq. (\ref{GMO2}) reduces to a trivial identity.

    The  l.h.s and r.h.s of  Eq. (\ref{GMO1})
 is $0.1867\,\rm GeV$ and  $0.1966\,\rm GeV$  respectively, showing a small 
$SU(3)$ symmetry breaking effects, of the order  $d=0.05$, the ratio of 
the difference between the l.h.s and r.h.s to the average 
of the two quantities and one would expect similar  amount of symmetry
breaking in hyperon semileptonic decays. 

  By making  a substitution
$\bar{u}s \to \bar{u}\gamma_{\mu}\gamma_{5}s$ in the $V-$spin relations
for the matrix elements of $\bar{u}s$ in Eqs. (\ref{su1},\ref{su2}),
the two GMO relations for the axial vector current matrix elements 
are obtained
\bea
&&\kern-1.5cm <\frac{1}{2}\Sigma^{0}+\frac{\sqrt{3}}{2}\Lambda|\bar{u}\gamma_{\mu}\gamma_{5}s|\Xi^{-}>\nonumber\\
&&= -<p|\bar{u}\gamma_{\mu}\gamma_{5}s|\frac{1}{2}\Sigma^{0} +\frac{\sqrt{3}}{2}\Lambda>\label{asu1}\\
&&\kern-1.5cm <\frac{\sqrt{3}}{2}\Sigma^{0}-\frac{1}{2}\Lambda|\bar{u}\gamma_{\mu}\gamma_{5}s|\Xi^{-}>\nonumber\\
&&= <p|\bar{u}\gamma_{\mu}\gamma_{5}s|\frac{\sqrt{3}}{2}\Sigma^{0} -\frac{1}{2}\Lambda>
\label{asu2}
\eea
    In terms of   $(g_{1}/f_{1})_{B\to B'}$ the axial
vector current to vector current form factor ratios~\cite{Bourquin}, the two GMO relations
become,
\bea
&&\kern-1.0cm (1/4)(g_{1}/f_{1})_{\Xi^{-}\to \Sigma^{0}} + (3/4)
(g_{1}/f_{1})_{\Xi^{-}\to \Lambda}\nonumber\\
&&= (1/4)(g_{1}/f_{1})_{\Sigma^{0}\to p} + (3/4)
(g_{1}/f_{1})_{\Lambda \to p}\label{GMOa}\\
&&\kern-1.0cm (3/4)[(g_{1}/f_{1})_{\Xi^{-}\to \Sigma^{0}} - 
(g_{1}/f_{1})_{\Xi^{-}\to \Lambda}]\nonumber\\
&&= -(3/4)[(g_{1}/f_{1})_{\Sigma^{0}\to p} - 
(g_{1}/f_{1})_{\Lambda \to p}]\label{GMOb}
\eea

    In the exact $SU(3)$ symmetry limit,  the l.h.s
 and r.h.s of  Eq. (\ref{GMOa})  as well as that of
 Eq. (\ref{GMOb}) are equal, given  by $F$ and $D$, respectively. 
Violation of the above relations comes from  first and second
order $SU(3)$ breaking terms, but second order terms could be less
important because of possible cancellation in $(g_{1}/f_{1})_{B\to B'}$.
Thus the validity of the above relations would depend essentially 
on  first order $SU(3)$ symmetry breaking effects.

   In   the presence of  $SU(3)$ symmetry breaking, the l.h.s and r.h.s
 of Eq. (\ref{GMOa}) and  Eq. (\ref{GMOb}) thus differ and are given by,
\bea
&& L_{1} = F + (1/4)\,d_{\Xi^{-}\to \Sigma^{0}} +
 (3/4)\,d_{\Xi^{-}\to\Lambda},\nonumber\\
&&R_{1} = F + (1/4)\,d_{\Sigma^{0}\to p} +
(3/4)\,d_{\Lambda \to p}
\label{GMOa1}
\eea
and 
\bea
&& L_{2} = D + (3/4)\,(d_{\Xi^{-}\to \Sigma^{0}}
 -d_{\Xi^{-}\to\Lambda}), \nonumber\\
&&  R_{2}= D - (3/4)\,(d_{\Sigma^{0}\to p} -
d_{\Lambda \to p}) 
\label{GMOb1}
\eea
with  $d_{B\to B^{'}}$ the corresponding symmetry breaking term.

    The differences $\Delta_{1}= L_{1} - R_{1}$ and
 $\Delta_{2}= L_{2} - R_{2}$ depend only on the symmetry breaking terms
 and are  measures of $SU(3)$ symmetry breaking. 

     From the measured values in Table
\ref{table1}, we have
\bea
&&\kern-1.0cm  L_{1} =0.490\pm 0.05, \quad R_{1} =0.453\pm 0.015, \nonumber\\
&& \Delta_{1}= 0.036\pm 0.065\label{R1}\\
&&\kern-1.0cm  L_{2} =0.720\pm 0.075, \quad R_{2} =0.793\pm 0.024,\nonumber\\
&& \Delta_{2}=-0.073\pm 0.10 \label{R2}
\eea
showing on average, an  amount of $SU(3)$ breaking  of $4\%$ from
 $\Delta_{1}$ and $10\%$ from
$\Delta_{2}$ (ignoring experimental errors), to be compared with
an amount of $SU(3)$ breaking of $5\%$ in   $<B^{'}|\bar{u}s|B>$ .

    The  $SU(3)$-symmetric fit in Ref.~\cite{Gensini} produces an  $SU(3)$
value $(g_{1}/f_{1})_{\Sigma^{-}\to n}=-0.3178$ to be compared with the
 measured value of $-0.340\pm 0.017$. This implies an $SU(3)$ breaking 
of $6.5\%$.  Taking $d_{\Sigma^-\to n}=-0.034$, we obtain
\be
F =  0.464 + 017, \quad D =  0.805  - 0.017 
\label{FD}
\ee
    Thus  $d_{\Sigma^{-}\to n}$
 makes  a rather small contribution to  $F$ and $D$. This
provides us with a quite precise determination of $F$ and $D$, using the
two GMO relations we found. With $F$ and $D$  determined, the symmetry 
breaking term $d_{B\to B^{'}} $ are then obtained from the 
measured  $(g_{1}/f_{1})_{B\to B'}$ as shown in Table 1.
    One would expect second order $SU(3)$ symmetry breaking
effects to be quite small for both vector and axial vector current
matrix elements.
   The determination of $V_{us}$ could be made by 
using the measured $g_{1}/f_{1}$ and neglecting second order
 $SU(3)$ symmetry breaking effects in the vector current form factor
$f_{1}$.    The values for $V_{us}$ thus obtained in 
Refs.~\cite{Gensini,Cabibbo2} are consistent with each other, and 
also agrees with the KTeV value from  neutral kaon semileptonic 
decays, as quoted in Ref.~\cite{Cabibbo2} : $V_{us} = 0.2250\pm 0.0027$\cite{Cabibbo2}, $V_{us} =0.22376\pm 0.00259$~\cite{Gensini} , to be compared with $V_{us} = 0.2252\pm 0.0005_{\rm KTeV}\pm 0.0009_{\rm ext}$ from KTeV.
    For a more precise determination of $V_{us}$
from hyperon decays one would need to compute second order 
$SU(3)$ symmetry breaking effects for $f_{1}$, as done in lattice 
calculation of the $K_{l3} $ $f_{+}(0)$ form factor which gives 
$V_{us} = 0.2246\pm 0.0012$~\cite{PDG}.
    Since $V_{us}$ obtained from hyperon semileptonic 
decays~\cite{Gensini,Cabibbo2} is close to the value  extracted
 from the measured average of  $V_{us}f_{+}(0)=0.2163$ of  various $K_{l3}$ 
measurements~\cite{Sciascia}, with corrections from  second order $SU(3)$ 
breaking for $f_{+}(0)$ given by chiral perturbation 
theory and lattice QCD calculations, second order $SU(3)$ breaking in 
the vector current form factor $f_{1}$  would be very
small, as mentioned above.
\vspace*{-0.25cm}
\section{Conclusion}
    The  GMO  relation for baryon mass difference is quite general
 and can be derived for the axial vector current matrix elements in
 hyperon semileptonic decays. This relation provides evidence for 
 an $SU(3)$ breaking effect of  the order $5-11\%$  in  hyperon
 semileptonic decays. The small symmetry breaking effect we find   also
confirms the success of the Cabibbo model for hyperon semileptonic decays.
    Finally, these GMO relations could be used as 
experimental constraints  on the $SU(3)$ symmetry breaking terms 
in theoretical calculations.
\vspace*{-0.25cm}
\section{Acknowledgments}
I would like to thank  S. Narison and 
the  organizers of QCD 2014 for the warm hospitality extended to me 
at Montpellier.

\end{document}